\documentclass[oupdraft]{bio}
\usepackage{url}
\usepackage{bm}
\newtheorem{Remark}{Remark}

\DeclareMathOperator*{\argmin}{arg\,min}

\usepackage{algorithm}
\usepackage[noend]{algpseudocode}
\usepackage[]{algpseudocode}
\usepackage{algorithmicx}
\usepackage{amsfonts}

\algnewcommand{\IfThenElse}[3]{% \IfThenElse{<if>}{<then>}{<else>}
  \State \algorithmicif\ #1\ \algorithmicthen\ #2\ \algorithmicelse\ #3}

\algloopdefx{NFor}[1]{\textbf{for all} #1 \textbf{do}}
\algnewcommand\And{\textbf{and}}
\algnewcommand\Or{\textbf{or }}

\usepackage{color}
\usepackage{soul} % So I can use \st for strikethrough

\usepackage{color}
\definecolor{eGreen}{rgb}{.057, .549,.065}

\history{Received January  1, ****;
revised January 1, ****;
accepted for publication January 1, ****}

\begin{document}

% Title of paper
\title{A constrained sparse additive model  for treatment effect-modifier selection}

% List of authors, with corresponding author marked by asterisk
\author{HYUNG G. PARK$^\ast$, EVA PETKOVA, THADDEUS TARPEY,\\
R. TODD OGDEN\\[4pt]
% Author addresses
\textit{Division of Biostatistics, Department of Population Health,
New York University,\\
180 Madison Ave., New York, USA} \\
\textit{Department of Biostatistics, Columbia University, 
168st, New York, 
USA}
\\[2pt]
% E-mail address for correspondence
{parkh15@nyu.edu}}

% Running headers of paper:
\markboth%
% First field is the short list of authors
{H. G. Park and others}
% Second field is the short title of the paper
{Sparse additive models for interactions}

\maketitle

% Add a footnote for the corresponding author if one has been
% identified in the author list
\footnotetext{To whom correspondence should be addressed.}

\begin{abstract}
{Sparse additive modeling is a class of effective methods for performing high-dimensional nonparametric regression. This paper develops a sparse additive model focused on estimation of treatment effect-modification with simultaneous treatment effect-modifier selection. We propose a version of the sparse additive model uniquely constrained to estimate the interaction effects between treatment and  pretreatment covariates, while leaving the main effects of the pretreatment covariates unspecified. The proposed regression model can effectively identify treatment effect-modifiers that exhibit possibly nonlinear interactions with the treatment variable, that are relevant for making optimal treatment decisions. A set of simulation experiments and an application to a dataset from a randomized clinical trial are presented to demonstrate the method. 
 }
{Individualized treatment rules;  Sparse additive models; Treatment effect-modifiers}
\end{abstract}

\section{Introduction}\label{sec.introduction}

Identification of patient characteristics influencing treatment responses, which are often termed treatment effect-modifiers or treatment effect-moderators, is a top research priority in precision medicine.  %\citep[e.g.,][]{Murphy, Robins}. 
In this paper, we develop a flexible yet simple and intuitive regression approach to identifying treatment effect-modifiers 
%identifying treatment effect-modifying  pretreatment patient characteristics 
from a potentially large number of pretreatment patient characteristics. % in a randomized clinical trial. 
In particular, we utilize a sparse additive model \citep{SAM} to 
conduct effective treatment effect-modifier selection.

The clinical motivation behind the development of a high-dimensional additive regression model 
that can handle a large number of pretreatment covariates, 
specifically designed to model treatment effect-modification  and variable selection,  
is a randomized clinical trial \citep{embarc} for treatment of major depressive disorder. 
A large number of baseline patient characterisitics were collected from each participant prior to randomization and treatment allocation. 
 The primary goal of this study  is to 
discover biosignatures of heterogeneous treatment response. %, which can be used for optimization of an individualized treatment rule for future patients. 
%As we move towards  optimizing individualized treatment rules, %and discovering the treatment implications from this vast amount of patient data, 
As we move towards  optimizing treatment decisions for individual patients, discovering  and identifying which pretreatment patient characteristics influence treatment effects  has the potential to significantly enhance clinical reasoning in practice \citep[see, e.g.,][]{Royston}.

The major challenge in efficiently modeling treatment effect-modification from a randomized clinical trial dataset is that variability due to treatment effect modification  (i.e., the treatment-by-covariates interaction effects on outcomes) is typically dwarfed by a relatively large degree of non-treatment-related variability (i.e., the main effects of the pretreatment covariates on treatment outcomes). 
 In particular, in regression, %in regression models,
  due to potential confounding between the main effect of the covariates and the  treatment-by-covariates interaction effect, misspecification of the covariate main effect 
may significantly influence  estimation of the treatment-by-covariates interaction effect.  
%with the result of a potentially severe estimation inconsistency to the true underlying interaction effect, unless an appropriate structure is imposed on the regression models. % \citep{CSIM}.   

A simple and elegant linear model-based approach to modeling the treatment-by-covariates interactions, termed the modified covariate (MC) method, 
that is robust against model misspecification of the covariates' main effects 
was developed by
 \cite{MC}. 
The method utilizes a simple parameterization and bypasses the need to model the  main effects of the covariates. 
 See also \cite{Murphy, LU.2011, A.learning.Shi2016, A.learning.Jeng2018}  
 for the similar  linear model-based  approaches to estimating the treatment-by-covariates interactions. 
%These methods are  robust against model misspecification of the covariates' main effect.   
 However, these % MC method  (as well as the other linear model-based approaches) 
%  linear model-based
   approaches assume a stringent linear model to specify the treatment-by-covariates interaction effects, 
 and are limited to the  binary-valued treatment variable case only. 
 In this work, we extend the optimization framework of \cite{MC} 
for modeling the treatment-by-covariates interactions  
using a more flexible regression setting  based on %generalized
 additive models \citep{GAM} that can accommodate more than two treatment conditions, 
 while allowing an unspecified main effect of the covariates. Additionally, via an appropriate regularization, the proposed approach  simultaneously achieves treatment effect-modifier selection in estimation.

%The paper is organized as follows. 
In Section~\ref{sec.models}, we introduce an additive model that has both the unspecified main effect of the covariates  and the treatment-by-covariates interaction effect additive components. Then we develop an optimization framework specifically targeting the interaction effect additive components of the model, with a sparsity-inducing regularization parameter to encourage sparsity in the set of component functions. In Section~\ref{sec.estimation}, we develop a coordinate decent algorithm to estimate the interaction effect part of the model and consider an optimization strategy for estimating individualized treatment rules. In Section~\ref{sec.simulation}, we illustrate the performance of the method in terms of treatment effect-modifier selection and estimation of optimal individualized treatment rules through simulation examples. Section~\ref{sec.application} provides an illustrative application from a depression clinical trial, and the paper concludes with discussion in Section~\ref{sec.conclusion}.

\section{Models} \label{sec.models} 

Let $A \in \{1,\ldots, L\}$ denote a treatment variable assigned with associated probabilities $(\pi_1,\ldots, \pi_L)$, $\sum_{a=1}^{L} \pi_a =1$, and $\pi_a > 0$, and let $\bm{X} = (X_1,\ldots, X_p)^\top \in \mathbb{R}^p$ denote pretreatment covariates, independent of $A$ (as in the case of a randomized trial). If the treatment assignment depends on the covariates, then the probabilities $\{ \pi_a\}_{a \in \{1,\ldots,L \}}$ can be replaced by propensity scores $\{ \pi_a(\bm{X}) \}_{a \in \{1,\ldots,L \}}$ 
 which would typically need to be estimated from data. 
 We let $Y^{(a)} \in \mathbb{R}$ (a = 1,\ldots,L) 
 be the potential outcome if the patient received treatment $A=a$ $(a = 1,\ldots,L)$; we only observe $Y = Y^{(A)}$, $A$ and $\bm{X}$. 
 
Throughout the paper, we assume that   $\mathbb{E}[Y|A = a] = 0 \ (a = 1,\ldots,L)$, without loss of generality, i.e., the main effect of $A$  on the outcome is centered at $0$. 
This  is only to suppress the treatment $a$-specific intercepts in regression models in  order to simplify the exposition, and  can be achieved by removing the treatment level $a$-specific means from $Y \in \mathbb{R}$. 
We model the treatment outcome $Y$  by  the following additive model:

%Suppose the treatment variable $A \in \mathcal{A}=  \{1,\ldots, L\}$ is assigned with associated probabilities $(\pi_1,\ldots, \pi_L)$, $\sum_{a=1}^{L} \pi_a =1$, and $\pi_a > 0$, independent of the pretreatment covariates $\bm{X} = (X_1,\ldots, X_p)^\top \in \mathbb{R}^p$, as in the case of RCTs.  Throughout,  if the treatment assignment depends on the covariates, then the probabilities $\{ \pi_a\}_{a \in \{1,\ldots,L \}}$ should be replaced by propensity scores $\{ \pi_a(\bm{X}) \}_{a \in \{1,\ldots,L \}}$  which would have to be estimated from data.  The treatment response variable $Y \in \mathbb{R}$ is modeled by  the following additive model: 

%\begin{equation} \label{the.model}
%\mathbb{E}[Y | \bm{X} , A=a]  =  \mu^\ast(\bm{X} ) +  \sum_{j=1}^p g_{j,a}^\ast(X_{j}) \quad (a=1,\ldots,L) 
%\end{equation}
\begin{equation} \label{the.model}
\mathbb{E}[Y | \bm{X} , A=a]  = \underbrace{\mu^\ast(\bm{X})}_{\bm{X} \mbox{ ``main'' effect}} +   \underbrace{ \sum_{j=1}^p g_{j,a}^\ast(X_{j})}_{A\mbox{-by-}\bm{X} \mbox{ interactions}}
 \quad (a=1,\ldots,L) 
\end{equation}
In model (\ref{the.model}), 
 the first term $\mu^\ast(\bm{X})$ %, an unspecified term, 
  does not depend on the treatment variable $A$ 
 and thus the $A$-by-$\bm{X}$ interaction effects are  
 determined only by the  second component
$\sum_{j=1}^p g_{j,A}^\ast(X_{j})$. 
In terms of modeling treatment effect-modification, 
the term $\sum_{j=1}^p g_{j,A}^\ast(X_{j})$ in (\ref{the.model}) corresponds to 
the ``signal'' component,  
whereas the term $\mu^\ast(\bm{X} )$ corresponds to  a ``nuisance'' component.

In (\ref{the.model}), 
for each individual covariate  $X_j$, 
we utilize a  treatment $a$-specific smooth $g_{j,a}^\ast$ separately for each treatment condition $a \in \{1,\ldots,L \}$. However, it is useful to treat the set of treatment-specific smooths for $X_j$ as a single unit, i.e., a single component function $g_j^\ast = \{ g_{j,a}^\ast \}_{a \in \{1,\ldots,L \}}$, for the purpose of treatment effect-modifier variable selection.

In model (\ref{the.model}), to separate $\mu^\ast(\bm{X})$ from the component  $\sum_{j=1}^p g_{j,a}^\ast(X_{j})$ and obtain an identifiable representation, 
without loss of generality, 
%we assume $\mathbb{E}[\mu^\ast(\bm{X})] =0$ and  $\mathbb{E}[Y | A=a] =0$ $(a=1,\ldots,L)$, i.e., the main effects of $\bm{X}$ and $A$ are centered at $0$, which can be achieved by regressing out the treatment level $a$-specific means.   Further, without loss of generality,  for model identifiability, 
 we assume that the set of  treatment $a$-specific smooths 
 $\{g^\ast_{j,a}\}_{a \in \{1,\ldots,L \}}$  of the $j$th component function $g_{j,A}^\ast(X_j)$ satisfies a condition:  
\begin{equation} \label{the.condition}
\mathbb{E}[g_{j,A}^\ast( X_j) | X_j ]  = \sum_{a=1}^L \pi_a g_{j,a}^\ast(X_j) = 0 \quad  
  \mbox{almost surely,} \quad (j=1,\ldots,p).
\end{equation}
%This condition indicates that, for each $j$, there are only $L-1$ arbitrary  unrestricted functions $g_{j,a}^\ast$ among the $L$ number of $A$-by-$X_j$ interaction effect smooths $\{g^\ast_{j,a}\}_{a \in \{1,\ldots,L \}}$  of the component function $g_j^\ast$. This implies that, for example, the smooth $g_{j,1}^\ast$ that corresponds to the treatment condition of $A=1$ is identified by  the other ($L-1$) functions:  $g_{j,1}^\ast(X_j) = - \pi_1^{-1}\sum_{a=2}^{L} \pi_a g_{j,a}^\ast(X_j)$ almost surely.  
Condition (\ref{the.condition}) 
implies 
$\mathbb{E}\big[ \sum_{j=1}^p g_{j,A}^\ast( X_j) | \bm{X} \big]  =0$ almost surely, 
and separates the  $A$-by-$\bm{X}$ interaction effect component,  
$\sum_{j=1}^p g_{j,a}^\ast(X_{j})$, from the $\bm{X}$ ``main'' effect component, $\mu^\ast(\bm{X})$, in model (\ref{the.model}). 
%In other words,  the identifiability constraint (\ref{the.condition})  prevents confounding between the main and interaction effect terms. 
For model (\ref{the.model}), we assume 
$Y = \mathbb{E}[Y |\bm{X}, A]  + \epsilon$, 
where $\epsilon$  is a zero-mean noise with finite variance, independent of $\bm{X}$ and $A$.

%\begin{Notation}
%For a single component $j$ and
% general (measurable) link functions $g_{j} = \{ g_{j,a} \}_{a \in \{1,\ldots,L \}}$, 
%define the $L^2$ norm of $g_j$ as 
%$\lVert g_j \rVert \ = \ \sqrt{\mathbb{E}\big[  g_{j,A}^2(X_j) 
% \big]}$, 
%where expectation is taken with respect to the joint distribution of $(A, X_j)$. (When there is no confusion, we also use $\lVert \cdot \rVert$ to denote the $L^2$ norm of a real vector.)
%For a set of random variables $(A, X_j)$ $(j=1,\ldots, p)$,  
%let $\mathcal{H}_j = \{ g_j  \mid  \mathbb{E}[  g_{j,A}(X_j)] =0,  \lVert g_j \rVert < \infty \}
%$ $(j=1,\ldots,p)$ 
%with inner product on the space defined as 
%$\langle g_j, f_{j} \rangle = \mathbb{E}[  g_{j,A}(X_j), f_{j,A}(X_j)]$. Sometimes we also write $g_j := g_{j,A}(X_j)$ for the notational simplicity. 
%Throughout the paper, vectors and matrices are denoted by bold-face letters, and estimates are denoted with a hat. 
%For example, we use $\hat{\bm{g}}_j \in \mathbb{R}^n$ to represent the vector of evaluations of the fitted functions $\hat{g}_j$ 
%at the $n$ observed values of $(X_{ij}, A_i)$ $(i=1,\ldots,n)$, 
%i.e., 
%$\hat{\bm{g}}_j = (\hat{g}_{j,A_1}(X_{1j}), \ldots, \hat{g}_{j,A_n}(X_{nj}) )^\top$. 
%\end{Notation}

%\begin*{Notation}
\textit{Notation}: 
%{\bf Notation} 
\textit{For a single component $j$ and
 general (measurable) component functions $g_j = \{ g_{j,a} \}_{a \in \{1,\ldots,L \}}$, 
define the $L^2$ norm of $g_j$ as 
$\lVert g_j \rVert \ = \ \sqrt{\mathbb{E}\big[  g_{j,A}^2(X_j) 
 \big]}$, 
where expectation is taken with respect to the joint distribution of $(A, X_j)$. (When there is no confusion, we also use $\lVert \cdot \rVert$ to denote the $L^2$ norm of a real vector.)
For a set of random variables $(A, X_j)$, %$(j=1,\ldots, p)$,  
let $\mathcal{H}_j = \{ g_j  \mid  \mathbb{E}[  g_{j,A}(X_j)] =0,  \lVert g_j \rVert < \infty \}
$ %$(j=1,\ldots,p)$ 
with inner product on the space defined as 
$\langle g_j, f_j \rangle = \mathbb{E}[  g_{j,A}(X_j), f_{j,A}(X_j)]$. Sometimes we also write $g_j := g_{j,A}(X_j)$ for the notational simplicity.}
%Throughout the paper, vectors and matrices are denoted by bold-face letters, and estimates are denoted with a hat. For example, we use $\hat{\bm{g}}_j \in \mathbb{R}^n$ to represent the vector of evaluations of the fitted functions $\hat{g}_j$ at the $n$ observed values of $(X_{ij}, A_i)$ $(i=1,\ldots,n)$, i.e., $\hat{\bm{g}} = (\hat{g}_{A_1}(X_{1}), \ldots, \hat{g}_{A_n}(X_{n}) )^\top$. 
%\end*{Notation}

 Under model  (\ref{the.model}) subject to (\ref{the.condition}), the component functions $\{ g_j^\ast, j=1,\ldots,p\}$ associated with the $A$-by-$\bm{X}$ interaction effect 
can be viewed 
as the solution to the %following
 constrained %least-squares
 optimization: 
 \begin{equation} \label{the.criterion}
\begin{aligned}
\{  g_{j}^\ast  \} 
\quad = \quad & \underset{ \{ g_{j}   \in \mathcal{H}_j \} }{\text{argmin}}
& &E \bigg[ \big\{ Y -  \mu^\ast( \bm{X})  -  \sum_{j=1}^p g_{j,A}(X_{j})    \big\}^2 \bigg]  \\
& \text{subject to} & & \mathbb{E}[g_{j,A}( X_j) | X_j ]  = 0 \quad (j=1,\ldots,p),  
\end{aligned}
\end{equation}
where $\mu^\ast(\bm{X})$ is given from the assumed model (\ref{the.model}) (and is considered as fixed in (\ref{the.criterion})). 
Since the minimization  
in (\ref{the.criterion})  is in terms of  
$\{ g_j \in \mathcal{H}_{j}, j=1,\ldots,p\}$,  
the objective 
function 
part of the right-hand side  
of (\ref{the.criterion}) 
can be reduced to: 
 \begin{equation*}\label{the.criterion2}
\begin{aligned}
& \argmin_{ \{ g_{j} \in \mathcal{H}_j \}  } \   \mathbb{E}\bigg[   Y^2  +  \big\{ \sum_{j=1}^p g_{j,A}(X_{j}) \big\}^2   - 2 \big\{\sum_{j=1}^p g_{j,A}(X_{j})\big\} Y + 2 \big\{\sum_{j=1}^p g_{j,A}(X_{j})\big\} \mu^\ast(\bm{X}) \bigg]  \\
  =&  \argmin_{ \{ g_{j} \in \mathcal{H}_j \} } \   \mathbb{E}\bigg[    Y^2  +  \big\{\sum_{j=1}^p g_{j,A}(X_{j})\big\}^2    - 2 \big\{\sum_{j=1}^p g_{j,A}(X_{j}) \big\} Y +  2 \mu^\ast(\bm{X}) \mathbb{E}\bigg[  \sum_{j=1}^p g_{j,A}(X_{j})  | \bm{X} \bigg] \bigg] \\ 
   =&  \argmin_{ \{ g_{j} \in \mathcal{H}_j \}}  \  \mathbb{E}\bigg[   Y^2  +  \big\{\sum_{j=1}^p g_{j,A}(X_{j}) \big\}^2    - 2 \big\{\sum_{j=1}^p g_{j,A}(X_{j})\big\} Y   \bigg],  
\end{aligned}
\end{equation*}
where 
the %first equality 
second line is  from an application of the iterated expectation rule conditioning on $\bm{X}$ and 
the %second equality
third line follows from the constraint $ \mathbb{E}[g_{j,A}( X_j) | X_j ]  = 0$ $(j=1,\ldots,p)$ on the right-hand side of (\ref{the.criterion}). 
Therefore, representation 
(\ref{the.criterion}) can be simplified to: 
 \begin{equation} \label{LS4}
\begin{aligned}
\{  g_{j}^\ast \} 
\quad = \quad & \underset{\{ g_{j}  \in \mathcal{H}_{j} \}}{\text{argmin}}
& &\mathbb{E} \bigg[ \big\{ Y -   \sum_{j=1}^p g_{j,A}(X_{j})    \big\}^2 \bigg]  \\
& \text{subject to} & & \mathbb{E}[g_{j,A}( X_j) | X_j ]  = 0 \quad (j=1,\ldots,p),  
\end{aligned}
\end{equation}

Representation (\ref{LS4}) of the  component functions $\{ g_j^\ast, j=1,\ldots,p\}$ of  the underlying model (\ref{the.model}) is particularly useful 
when the  (high-dimensional) ``nuisance'' function  $\mu^\ast$ in (\ref{the.model}) is complicated and prone to specification error. 
Note, 
\begin{equation}\label{orthogonality}
\mathbb{E} \bigg[  \mu^\ast( \bm{X} ) \sum_{j=1}^p g_{j,A} \big( X_j \big)  \bigg]=
\mathbb{E} \bigg[ \mu^\ast( \bm{X} )  \sum_{j=1}^p   \mathbb{E} \big[  g_{j,A} \big( X_j \big)   | \bm{X} \big]\bigg] =0, 
\end{equation} 
indicating that the component $ \sum_{j=1}^p g_{j,A} \big( X_j \big) \in \mathcal{H}_1 + \cdots + \mathcal{H}_p$ (subject to the identifiability constraint $\mathbb{E}[g_{j,A}( X_j) | X_j ]  = 0$)  
is structured to be orthogonal to the $\bm{X}$ main effect $\mu^\ast(\bm{X})$. This orthogonality property is useful for estimating the additive model $A$-by-$\bm{X}$ interaction effect $\sum_{j=1}^p g_{j,A}^\ast(X_{j})$, in the presence of the unspecified $\mu^\ast(\bm{X})$ of model (\ref{the.model}).

 Under model (\ref{the.model}), 
  the potential treatment effect-modifiers among $\{ X_j, j=1,\ldots,p  \}$   
   enter the model only through the interaction effect term $\sum_{j=1}^p g_{j,A}^\ast(X_j)$ that associates the treatment $A$ to the treatment outcome $Y$. 
  \cite{SAM} proposed sparse additive modeling (SAM) for component selection in  high-dimensional additive models  with a large $p$.  
 As in SAM, to deal with a large $p$ and to achieve treatment effect-modifier selection, we  impose  sparsity  
 on the set of  component functions $\{ g_j^\ast, j=1,\ldots,p\}$ associated with the interaction effect term of model (\ref{the.model}), 
   under the often practical and reasonable assumption that most covariates are irrelevant as treatment effect-modifiers.   
This sparsity structure on the index set $\{ j\}$ for the nonzero component functions   $\{  g_{j}^\ast :  g_j^\ast \ne 0 \} $ 
can be usefully incorporated into 
the optimization-based criterion (\ref{LS4}) in representing $\{  g_{j}^\ast \}$:  
 \begin{equation} \label{LS5}
\begin{aligned}
\{  g_{j}^\ast \} 
\quad = \quad & \underset{\{ g_{j}  \in \mathcal{H}_{j} \}}{\text{argmin}}
& &\mathbb{E} \bigg[ \big\{ Y -   \sum_{j=1}^p g_{j,A}(X_{j})    \big\}^2 \bigg]   \  + \ \lambda \sum_{j=1}^p \lVert g_{j} \rVert \\
& \text{subject to} & & \mathbb{E}[g_{j,A}( X_j) | X_j ]  = 0 \quad (j=1,\ldots,p),  
\end{aligned}
\end{equation}
for a sparsity-inducing parameter $\lambda  \ge 0$. 
The term $\sum_{j=1}^p \lVert g_{j} \rVert$ in (\ref{LS5})  behaves like an $L^1$ ball across different components $\{ g_j, j=1,\ldots,p\}$ to encourage sparsity in the set of component functions. 
For example, a large value of  $\lambda$ on the right-hand side of (\ref{LS5})  will generate a sparse solution with many component functions $g_j^\ast$  on the left-hand side set exactly to zero.

  \section{Estimation} \label{sec.estimation}

\subsection{Model estimation}

For each $j$, 
the minimizer $g_j^\ast$ of the optimization problem 
 (\ref{LS5}) has  a component-wise closed-form expression.

\begin{theorem} \label{theorem1}
Given $\lambda \ge 0$, 
 the minimizer  $g_j^\ast \in \mathcal{H}_j$ 
of (\ref{LS5})  satisfies: 
\begin{equation}\label{g.solution}
g_{j,A}^\ast(X_j ) \ =\  \left[ 1- \frac{\lambda}{\lVert f_{j} \rVert} \right]_{+}  f_{j,A}(X_j)  \quad \mbox{almost surely},  
\end{equation} 
where 
\begin{equation}\label{proj.1}
f_{j,A}(X_j)  \ =  \ \mathbb{E}[R_{j} |  X_j, A ] \ -  \ \mathbb{E}[R_{j} | X_j ],  
\end{equation}
%corresponding  to the projection of $R_j$ onto  $\mathcal{H}_j$ subject to the constraint in (\ref{LS5}), 
in which 
\begin{equation}\label{partial.residual}
R_j = Y - \sum_{ j^\prime \ne j}g_{j^\prime,A}^\ast( X_{j^\prime} ) 
 \end{equation}
represents  the $j$th partial residual. 
 In (\ref{g.solution}), $Z_{+} = \max(0,Z)$ represents the positive part of $Z$. 
\end{theorem}
Note that the $f_{j,A}(X_j)$ correspond to the projections of $R_j$ onto 
 $\mathcal{H}_j$ subject to the constraint in (\ref{LS5}).
The proof of Theorem~\ref{theorem1} is in the Supplementary Materials.

The component-wise expression (\ref{g.solution}) for $g_j^\ast$ suggests that 
we can employ a coordinate descent algorithm \citep[e.g.,][]{Tseng2001} to solve (\ref{LS5}). 
Given a sparsity parameter 
 $\lambda \ge 0$, 
we can use a standard backfitting algorithm used in fitting additive models (\cite{GAM}) 
that fixes the set of current approximates for $g_{j'}^\ast$ at all $j' \ne  j$,  
and obtains a new approximate of $g_{j}^\ast$ by equation (\ref{g.solution}), 
and iterates through all $j$ until convergence. 
A sample version of the algorithm can be obtained by inserting sample estimates into the population expressions (\ref{partial.residual}), (\ref{proj.1}) and (\ref{g.solution}) for each coordinate $j$, 
which we briefly describe next.

Given  data $(X_{ij}, A_{i})$  $(i=1,\ldots,n; j=1,\dots, p)$, 
for each $j$,  let $\hat{R}_{j} = Y - \sum_{ j^\prime \ne j}\hat{g}_{j^\prime,A}^\ast( X_{j^\prime} )$, 
corresponding to the  data-version of the $j$th partial residual $R_j$ in (\ref{partial.residual}), 
where $\hat{g}_{j^\prime}^\ast$ represents a current estimate for $g_{j^\prime}^\ast$. 
%The function $g_j^\ast$ in (\ref{g.solution}) is estimated first by obtaining an estimate of  $f_j$ in  (\ref{proj.1}) and thensoft-thresholding $f_j$ by the shrinkage factor $\left[ 1- \frac{\lambda}{\lVert f_{j} \rVert} \right]_{+}$   in (\ref{g.solution}).
%For each $j$, 
We estimate $g_j^\ast$ in (\ref{g.solution}) in two steps: 
1)  estimate the function $f_j$ in 
 (\ref{proj.1});  
2) 
plug the estimate of $f_j$ into $\left[ 1- \frac{\lambda}{\lVert f_{j} \rVert} \right]_{+}$  
 in (\ref{g.solution}),  
to obtain the soft-thresholded estimate $\hat{g}_j^\ast$.

Although any linear smoothers can be utilized to obtain the estimators $\{ \hat{g}_j^\ast \}$ as described in Remark~\ref{remark1} at the end of this section, in this paper, 
we shall focus on regression spline-type estimators
which are particularly simple and computationally efficient to implement. 
Specifically, for each $j$, 
 the function 
 $g_j \in \mathcal{H}_j$ 
on the right-hand side of  (\ref{LS5}) will be represented by: 
\begin{equation} \label{eq.5}
g_{j,a}(X_j)  = \bm{\Psi}_j(X_j)^\top \bm{\theta}_{j,a} \quad (a=1,\ldots,L)
\end{equation}
for some prespecified $d_j$-dimensional basis $\bm{\Psi}_j(\cdot) \in \mathbb{R}^{d_j}$ (e.g., $B$-spline basis on evenly spaced knots on a bounded range for $X_j$) and a set of unknown treatment $a$-specific basis coefficients $\{ \bm{\theta}_{j,a}  \in \mathbb{R}^{d_j} \}_{a \in \{1,\ldots,L \}}$. 
Given representation (\ref{eq.5}) for the component function $g_j$, 
the constraint  
$\mathbb{E}[g_{j,A}(X_j) | X_j] =0$ in (\ref{LS5}) 
can be simplified to 
$\mathbb{E}[ \bm{\theta}_{j,A}  ] = \sum_{a=1}^{L} \pi_a \bm{\theta}_{j,a} = \bm{0}$. %, where $\bm{0}  \in \mathbb{R}^{d_j}$ denotes the vector of zeros. 
This constraint can be written succinctly in a matrix form as
\begin{equation}\label{lin.constr}
\bm{\pi}^{(j)} \bm{\theta}_j  = \bm{0},   
\end{equation} 
where $\bm{\theta}_j := (\bm{\theta}_{j,1}^\top, \bm{\theta}_{j,2}^\top, \ldots, \bm{\theta}_{j,L}^\top )^\top \in \mathbb{R}^{d_jL}$ is the vectorized version of the basis coefficients 
$\{\bm{\theta}_{j,a}\}_{a \in \{1,\ldots,L \}}$ in (\ref{eq.5}), 
and the $d_j \times d_jL$ matrix  $\bm{\pi}^{(j)} := \left(\pi_1 \bm{I}_{d_j}; \pi_2 \bm{I}_{d_j};  \ldots; \pi_L \bm{I}_{d_j}\right)$ where $\bm{I}_{d_j}$ denotes  the $d_j \times d_j$ identity matrix.

The restriction of the function  $g_j$ to the form (\ref{eq.5})  restricts also the minimizer $g_j^\ast$  in (\ref{g.solution}) (note, $g_{j,A}^\ast(X_j) = s_j^{(\lambda)} f_{j,A}(X_j)$, where $s_j^{(\lambda)} = \left[ 1- \lambda/ \lVert f_{j} \rVert  \right]_{+}$) to the form (\ref{eq.5}). In particular, 
we can express the function $f_j$ in (\ref{proj.1}) as:
 \begin{equation} \label{eq.4}
\begin{aligned}
 f_{j,A}( X_j)  
&=    \mathbb{E}[R_j | X_j, A ] -   \sum_{a=1}^L \pi_a  \mathbb{E}[R_{j} |  X_j, A=a ]  \\ 
  %&=    \bm{\Psi}_j(X_j)     \big( \bm{\theta}_{j,A}^\ast  -  \sum_{a=1}^L \pi_a  \bm{\theta}_{j,a}^\ast   \big) 
  &=    \bm{\Psi}_j(X_j)  \bm{\theta}_{j,A}^\ast  -    \bm{\Psi}_j(X_j)  \big(  \sum_{a=1}^L \pi_a  \bm{\theta}_{j,a}^\ast   \big) 
\end{aligned}
\end{equation} 
where $\{ \bm{\theta}_{j,a}^\ast\}_{a \in \{1,\ldots,L \}}   := \underset{\{ \bm{\theta}_{j,a} \in \mathbb{R}^{d_j} \}}{\text{argmin}}    \mathbb{E}\big[ \{R_j -  \bm{\Psi}_j(X_j)^\top   \bm{\theta}_{j,A} \}^2\big]$. 
The first term  $ \mathbb{E}[R_j | X_j, A ]$ in (\ref{eq.4}) 
corresponds  to the  $L^2$ projection of the $j$th partial residual $R_j$ onto the class of functions of the form (\ref{eq.5}) (without the imposition of the constraint (\ref{lin.constr})), whereas the second term  $-  \sum_{a=1}^L \pi_a  \mathbb{E}[R_{j} |  X_j, A=a ] $ centers the first term to satisfy the linear constraint  (\ref{lin.constr}). 
It is straightforward to verify that $f_j$, as given in  (\ref{eq.4}), corresponds to the $L^2$ projection of $R_j$ 
 onto the subspace of  measurable functions of the form (\ref{eq.5}) subject to the linear constraint (\ref{lin.constr}).

Let the $n \times d_j$ matrices 
 $\bm{D}_{j,a}$ $(a=1,\ldots,L)$ denote the %(treatment $a$-specific) 
 evaluation matrices  of the basis function $\bm{\Psi}_j(\cdot)$ on $X_{ij}$ $(i=1,\ldots,n)$ specific to the treatment $A=a$ $(a=1,\ldots,L)$, whose $i$th row is  the $1 \times d_j$ vector $\bm{\Psi}_j(X_{ij})^\top$  if  $A_i = a$, and a row of zeros $\bm{0}^\top$ if $A_i \ne a$. 
Then the column-wise concatenation 
of the design matrices $\{ \bm{D}_{j,a}\}_{a \in \{1,\ldots,L \}}$, i.e., the $n \times d_jL $ matrix
 $\bm{D}_j  = (\bm{D}_{j,1}; \bm{D}_{j,2}; \ldots; \bm{D}_{j,L})$,  
 defines the model matrix associated with the vectorized model coefficient  $\bm{\theta}_j \in \mathbb{R}^{d_jL}$, vectorized across 
$\{\bm{\theta}_{j,a}\}_{a \in \{1,\ldots,L \}}$ in representation (\ref{eq.5}). 
Then we can represent the function $g_{j,A}(X_j)$ in (\ref{eq.5}), based on the sample data, by the length-$n$ vector: 
%It follows that the function $g_{j,A}(X_j)$ in (\ref{eq.5}), based on the sample, can be represented by the length-$n$ vector: 
\begin{equation} \label{gj.vector1}
\bm{g}_j  = \bm{D}_j \bm{\theta}_j 
\end{equation}
subject to the linear constraint (\ref{lin.constr}).

When computing the data version of  the function $f_j$ in (\ref{proj.1}) which corresponds to
 the projection of $R_j$ onto the functional class of (\ref{eq.5}) subject to (\ref{lin.constr}),  
the linear constraint (\ref{lin.constr}) 
on $\bm{\theta}_j$ can be conveniently absorbed into the model matrix $\bm{D}_j$ in (\ref{gj.vector1}) 
by reparametrization, as we describe next. 
We can find a $d_jL \times d_j(L-1)$ basis matrix $\bm{Z}^{(j)}$, 
such that if we set  $\bm{\theta}_j = \bm{Z}^{(j)} \tilde{\bm{\theta}}_j$  for any arbitrary vector $\tilde{\bm{\theta}}_j \in \mathbb{R}^{d_j(L-1)}$, then the vector $\bm{\theta}_j \in \mathbb{R}^{d_jL}$ automatically satisfies the constraint (\ref{lin.constr}) 
$\bm{\pi}^{(j)} \bm{\theta}_j  = \bm{0}$. %,   satisfying  the constraint (\ref{lin.constr}). 
Such a basis matrix $\bm{Z}^{(j)}$, that  spans  the null space of the linear constraint (\ref{lin.constr}), 
can be constructed by a QR decomposition of the matrix 
$\bm{\pi}^{(j)\top}$. 
Then representation (\ref{gj.vector1}) can be reparametrized, 
in terms of the unconstrained vector $\tilde{\bm{\theta}}_j \in \mathbb{R}^{d_j(L-1)}$,  
by replacing $\bm{D}_j$ in (\ref{gj.vector1}) with the reparametrized model matrix $\tilde{\bm{D}}_j =  \bm{D}_j  \bm{Z}^{(j)}$: 
\begin{equation} \label{gj.vector2}
\bm{g}_j = \tilde{\bm{D}}_j \tilde{\bm{\theta}}_j. 
\end{equation}

Theorem~\ref{theorem1} indicates that 
the coordinate-wise minimizer  $g_j^\ast$ of (\ref{LS5}) can be estimated based on the sample by  
\begin{equation} \label{ghat.solution}
\hat{\bm{g}}_j^\ast  = \left[ 1- \frac{\lambda}{\sqrt{ \frac{1}{n} \lVert  \hat{\bm{f}}_j \rVert^2} } \right]_{+}  \hat{\bm{f}}_j 
\end{equation}
where 
\begin{equation} \label{f.solution}
\hat{\bm{f}}_j  = \tilde{\bm{D}}_j (\tilde{\bm{D}}_j^{\top} \tilde{\bm{D}}_j)^{-1} \tilde{\bm{D}}_j^{\top} \hat{\bm{R}}_j 
\end{equation}
in which $\hat{\bm{R}}_j  = \bm{Y} -  \sum_{j^\prime \ne j} \hat{\bm{g}}_{j^\prime}^\ast$ is the estimated $j$th partial residual vector. 
In (\ref{ghat.solution}), 
the norm $\lVert f_j \rVert $ 
of (\ref{g.solution}) is estimated 
by the vector norm $\sqrt{ \frac{1}{n} \lVert  \hat{\bm{f}}_j \rVert^2}$, and the shrinkage factor $(s_j^{(\lambda)} =) \left[ 1- \frac{\lambda}{\lVert f_{j} \rVert} \right]_{+} $ of (\ref{g.solution})
 is estimated by $(\hat{s}_j^{(\lambda)} =) \left[ 1- \frac{\lambda}{\sqrt{ \frac{1}{n} \lVert  \hat{\bm{f}}_j \rVert^2} } \right]_{+}$.

Based on the sample counterpart (\ref{ghat.solution}) of the 
coordinate-wise solution (\ref{g.solution}), a highly efficient coordinate descent algorithm can be conducted 
to simultaneously estimate all the component functions 
$\{g_j^\ast, j=1,\ldots,p\}$ in  (\ref{LS5}). 
At  convergence of the coordinate descent, 
we have a basis coefficient estimate associated with the representation (\ref{gj.vector2}), 
\begin{equation} \label{coef.hat}
 \hat{\tilde{\bm{\theta}}}_j =  \hat{s}_j^{(\lambda)}  (\tilde{\bm{D}}_j^{\top} \tilde{\bm{D}}_j)^{-1} \tilde{\bm{D}}_j^{\top} \hat{\bm{R}}_j
\end{equation}
which in turn implies an estimate 
$$\hat{\bm{\theta}}_j = (\hat{\bm{\theta}}_{j,1}^\top, \hat{\bm{\theta}}_{j,2}^\top, \ldots, \hat{\bm{\theta}}_{j,L}^\top   )^\top = \bm{Z}^{(j)} \hat{\tilde{\bm{\theta}}}_j$$ 
for the basis coefficient associated with  the representation (\ref{gj.vector1}). 
This gives an estimate of the  treatment $a$-specific function $g_{j,a}^\ast(\cdot)$ $(a=1,\ldots,L)$  
in model (\ref{the.model}): 
\begin{equation} \label{g.hat}
\hat{g}_{j,a}^\ast(\cdot) = \bm{\Psi}_j(\cdot)^\top  \hat{\bm{\theta}}_{j,a} \quad (a=1,\ldots,L)
\end{equation} 
 estimated within the class of functions (\ref{eq.5}) for a given tuning parameter  $\lambda$, 
  which controls the shrinkage factor $\hat{s}_j^{(\lambda)}$  in (\ref{coef.hat}). 
We summarize the computational procedure for the coordinate descent 
in Algorithm~\ref{algorithm1}.

\begin{algorithm}[H]
\caption{Coordinate descent } \label{algorithm1}
\begin{algorithmic} [1]
  \State  \textbf{Input}: Data $\bm{X} \in \mathbb{R}^n \times \mathbb{R}^p$, $\bm{A} \in \mathbb{R}^n$, $\bm{Y} \in \mathbb{R}^n$, and tuning parameter $\lambda$. 
   \State  \textbf{Output}:  Fitted functions $\{ \hat{\bm{g}}_j^\ast, j=1,\ldots,p  \}$. 
   \State Initialize $\hat{\bm{g}}_j^\ast = \bm{0}$ $\forall j$;  
   pre-compute the smoother matrices $ \tilde{\bm{D}}_j (\tilde{\bm{D}}_j^{\top} \tilde{\bm{D}}_j)^{-1} \tilde{\bm{D}}_j^{\top} $ in (\ref{f.solution}) $\forall j$. 
  \While 
    {until convergence of 
    $\{ \hat{\bm{g}}_j^\ast, j=1,\ldots,p  \}$,
    } {iterate through $j=1,\ldots,p$:} 
%   \For {Cycle through $j=1,\ldots,p$ until convergence of $\{ \hat{\bm{g}}_j^\ast, j=1,\ldots,p  \}$,}
%  \For {$j = 1,\ldots, p,$} 
    \State  Compute the partial residual $\hat{\bm{R}}_j  = \bm{Y} -  \sum_{j^\prime \ne j} \hat{\bm{g}}_{j^\prime}^\ast$ 
     \State  Compute $\hat{\bm{f}}_j$ in (\ref{f.solution}); 
     then compute the thresholded estimate $\hat{\bm{g}}_j^\ast$ in (\ref{ghat.solution}). 
  %   \EndFor
     \EndWhile
\end{algorithmic}
\end{algorithm}

  In Algorithm~\ref{algorithm1},  
 the projection matrices $ \tilde{\bm{D}}_j (\tilde{\bm{D}}_j^{\top} \tilde{\bm{D}}_j)^{-1} \tilde{\bm{D}}_j^{\top} $   $(j=1,\ldots, p)$  
only need to be computed once and therefore the coordinate descent can be performed efficiently.  
  In (\ref{ghat.solution}), 
  if the shrinkage factor   $\hat{s}_j^{(\lambda)} = \left[ 1- \frac{\lambda \sqrt{n}}{  \lVert  \hat{\bm{f}}_j \rVert } \right]_{+} = 0$,  
 the associated $j$th  covariate is absent from the model. 
The tuning parameter $\lambda \ge 0$ for treatment effect-modifier selection can be chosen to minimize
an estimate of the expected squared error of the fitted models, $\mathbb{E} \big[ \{ Y -   \sum_{j=1}^p \hat{g}_{j,A}^\ast(X_{j})\}^2 \big]$, over a dense grid of $\lambda$'s,  
estimated, for example, by cross-validation. Alternatively, one can utilize the network information criterion \citep[NIC;][]{NIC} which is a generalization of  the Akaike information criterion  %\citep[AIC;][]{AIC} 
in approximating the prediction error, 
for the case where the true underling model, i.e., model (\ref{the.model}), is not necessarily in the class of candidate models.  
Throughout the paper,   $\lambda$  is chosen to minimize $10$-fold cross-validated prediction error of the fitted models.

\begin{Remark} \label{remark1}
For coordinate descent, 
 any linear smoothers %for nonparametric regression 
 can be utilized to obtain 
 the sample counterpart (\ref{ghat.solution}) of the 
coordinate-wise solution (\ref{g.solution}), i.e., the method is not restricted to regression splines. 
To estimate the function $f_j$ in (\ref{proj.1}), 
we can estimate the first term 
$\mathbb{E}[R_{j} | X_j, A=a]$ 
 in   (\ref{proj.1}), using a 1-dimensional nonparametric smoother 
for each treatment level $a \in \{1,\ldots,L \}$ separately, 
based on the data $(\hat{R}_{ij}, X_{ij})$ $(i \in \{i : A_i = a\})$ corresponding  to the  data from the $a$th treatment condition; 
we can also estimate 
the second term $- \mathbb{E}[R_{j} | X_j]$ in  (\ref{proj.1}) based on the data 
 $(\hat{R}_{ij}, X_{ij})_{i \in \{1,\ldots,n\}}$    
 which corresponds to the set of data from all treatment conditions, 
 using a 1-dimensional nonparametric smoother. 
 Adding these two estimates evaluated at the $n$ observed values of $(X_{ij}, A_i)$ $(i=1,\ldots,n)$
 gives an estimate $\hat{\bm{f}}_j$ in (\ref{ghat.solution}). 
 Then we can compute the associated estimate $\hat{\bm{g}}_j^\ast$, which allows implementation of 
 the coordinate descent in Algorithm~\ref{algorithm1}. 
\end{Remark}

\subsection{Individualized treatment rule estimation}

 For a single time decision point,  
an individualized treatment decision rule (ITR), which we denote by  
 $\mathcal{D}(\bm{X}): \mathbb{R}^p \mapsto  \{1, \ldots, L\}$,   
 maps an individual with pretreatment characteristics 
$\bm{X}$ to one of the $L$ treatment options. 
% for a treatment recommendation. 
One natural measure 
for the effectiveness of an ITR $\mathcal{D}$ in precision medicine is the so-called  ``value'' ($V$) function \citep{Murphy2005}: 
\begin{equation} \label{value.eq}
V(\mathcal{D}) = \mathbb{E}[ \mathbb{E}[ Y | \bm{X}, A= \mathcal{D}(\bm{X})] ], 
\end{equation} 
which is the expected treatment response under a given ITR 
$\mathcal{D}$. If we assume that a larger value of $Y$ is better (without loss of generality), then 
the optimal ITR 
$\mathcal{D}$, which we write as $\mathcal{D}^{opt}$, can be naturally defined to be 
the rule that maximizes the value $V(\mathcal{D})$ (\ref{value.eq}). 
Such an optimal rule 
$\mathcal{D}^{opt}$ satisfies: 
% can be easily shown to satisfy: 
\begin{equation} \label{d.opt}
\mathcal{D}^{opt}(\bm{X}) =  \operatorname*{arg\,max}_{a \in \{1,\ldots,L \}} \   \mathbb{E}[ Y | \bm{X}, A=a ] \quad \mbox{(almost surely)}. 
\end{equation}

Much work has been carried out to develop methods to estimate the optimal ITR (\ref{d.opt}) using data from randomized clinical trials.  Machine learning approaches to estimating  (\ref{d.opt}), 
including the outcome weighted learning \citep[e.g.,][]{Zhao.2012, Zhao.2015, Song.2015} based on support vector machines (SVMs), tree-based classification \citep[e.g.,][]{Laber.Zhao.2015}, and adaptive boosting \citep{KANG.2014}, among others,  
 are often framed in the context of a (weighted) classification problem \citep{Zhang.classification, Zhao.2019}, 
where $\mathcal{D}^{opt}(\bm{X}) $ in (\ref{d.opt})  
 is regarded as  the optimal classification rule  
  for treatment with respect to the objective function (\ref{value.eq}). 
  
   Under model (\ref{the.model}),  
 $\mathcal{D}^{opt}(\bm{X}) $ in (\ref{d.opt}) 
is:  
$\mathcal{D}^{opt}(\bm{X}) =  \operatorname*{arg\,max}_{a \in \{1,\ldots,L \} }  \sum_{j=1}^p g_{j,a}^\ast(X_j)$, 
which can be estimated by:  
$\hat{\mathcal{D}}^{opt}(\bm{X}) =   \operatorname*{arg\,max}_{a \in \{1,\ldots,L \} }  \sum_{j=1}^p \hat{g}_{j,a}^\ast(X_j)$,  
where  $\hat{g}_{j,a}^\ast(\cdot)$ is given in (\ref{g.hat}) 
 at the convergence of the Algorithm~\ref{algorithm1}.  
The proposed approach (\ref{LS5}), which can be viewed as a regression-based approach 
 to estimate (\ref{d.opt}), 
 %The approach
approximates the conditional expectations $\mathbb{E}]Y | \bm{X}, A=a]$ $(a=1,\ldots,L)$ 
 based on the additive model (\ref{the.model}),  while maintaining robustness with respect to model misspecification of the 
 ``nuisance''  function 
 $\mu^\ast(\cdot)$ in (\ref{the.model}) 
which is not relevant in estimating $\mathcal{D}^{opt}(\bm{X})$ in (\ref{d.opt}).    
We illustrate the performance of  the estimator $\hat{\mathcal{D}}^{opt}$ %of $\mathcal{D}^{opt}(\cdot)$ 
 with respect to the value function  (\ref{value.eq}), through a set of simulation studies in Section~\ref{sec.sim.ITR}.

\subsection{Feature selection and transformation for individualized treatment rules}

Although machine learning approaches that attempt to directly maximize (\ref{value.eq}) 
without assuming some specific structure on $\mathbb{E}]Y | \bm{X}, A=a]$ $(a=1,\ldots,L)$ 
(unlike most of the regression-based approaches) 
are highly appealing, 
common machine learning approaches used in optimizing ITRs, including SVMs utilized in the outcome weighted learning, are often  hard to scale to large datasets, due to their taxing computational time.  
In particular, SVMs are viewed as ``shallow'' approaches (as opposed to a ``deep'' learning method that utilizes a learning model with many representational layers)  
and successful applications of  SVMs  often require first extracting useful representations for their input data manually 
or through some data-driven feature transformation (a step called feature engineering) \citep[see, e.g.,][]{feature.engineering}  %  to find an appropriate representation of data, 
to have more discriminatory power.  
%or some data-driven feature transformation. 
%The new features may not have the same interpretation as the original features, but they may have more discriminatory power in a different space than the original space.  
Generally, selection and transformation of relevant features can increase the  performance, scale and speed of a machine learning procedure. % for optimizing ITRs.  

As an added value, the proposed regression (\ref{LS5}) based on model (\ref{the.model}) provides  a practical %and efficient 
feature selection and transformation learning technique for optimizing ITRs.  
%We can utilize 
The set of component functions $\{g_j^\ast, j=1,\ldots,p \}$ in (\ref{LS5}) 
 can be used to define %a set of
  data-driven feature transformation functions for the original features $\{X_j, j=1,\ldots,p\}$. 
The resulting transformed features can be used as an input to a machine learning algorithm for optimizing  ITRs. % optimization.  
%for example, the outcome weighted learning,  for optimizing ITRs. 

In particular, we note that for each $j$, the component function $g_j^\ast$ in (\ref{LS5}) is defined independently from the $X_j$ main effect function $\mu^\ast$ in (\ref{the.model}),  
and therefore the corresponding transformed feature variable $g_{j,1}^\ast(X_j)$, which represents the $j$th feature $X_j$ in the new space, highlights only the ``signal'' nonlinearity in $X_j$ related to the $A$-by-$X_j$  
interactions  that is relevant to estimating $\mathcal{D}^{opt}$, 
while excluding the nonlinearity in $X_j$ related to the $X_j$ main effect which is irrelevant to the  ITR development. 
This ``de-noising''  procedure for each variable $X_j$ can be often very appealing, since irrelevant or partially relevant features can negatively impact the performance of a machine learning algorithm. 
Moreover, 
 a relatively large value of the tuning parameter $\lambda > 0$ in (\ref{LS5})  
would imply a set of sparse component functions $\{ g_j ^\ast\}$, 
  providing a means of feature selection for ITRs. 
For the most common case of  $L=2$ (binary treatment), we have 
$g_{j,2}^\ast(X_j) = - \pi_2^{-1} \pi_1 g_{j,1}^\ast(X_j)$ 
implied by the constraint  (\ref{the.condition}) that we impose, 
 which is simply a scalar-scaling of the function $g_{j,1}^\ast(X_j)$; this implies that, 
for each $j$, 
the mapping 
$X_j \mapsto g_{j,1}^\ast(X_j)$ specifies the feature transformation of $X_j$. 
We demonstrate the utility of this feature selection/transformation, 
%resulting in the transformed features $\{g_{j,1}^\ast(X_j) \}$, 
which we use as an input to the outcome weighted learning approach to optimizing ITRs,   
through a set of simulation studies in Section~\ref{sec.sim.ITR} and a real data application in Section~\ref{sec.application}.

\section{Simulation study}\label{sec.simulation}

\subsection{Treatment effect-modifier selection performance} \label{variable.selection.sim}

In this section, we will report simulation results illustrating the performance of  the treatment effect-modifier selection. 
The complexity of the model for studying the $A$-by-$\bm{X}$ interactions can be summarized in terms of the size of the index set for the component functions $\{g_{j}^\ast, j=1,\ldots,p \}$  that are not identically zero. We can ascertain the performance of a treatment effect-modifier selection method in terms of these component functions correctly or incorrectly estimated as nonzero.  
%The complexity of the model for studying the $A$-by-$\bm{X}$ interactions can be summarized in terms of thesize of the index set for the component functions $\{g_{j}^\ast \}_{j=1,\ldots,p}$  that are not identically zero,  which can be either correctly or incorrectly estimated as nonzeros.  
 To generate the data,  we use the following model: 
  \begin{equation} \label{sim3.model}
Y  = \sum_{j=1}^{10}  \cos(X_{j})  \ + \   (A-1.5)  X_1  + \  2 (A-1.5) \cos(X_2)  \  + \epsilon  \quad A  \in \{1,2\}, 
\end{equation} 
where $X_j$ $(j=1,\ldots,p)$,  $p \in \{50, 100\}$, is generated from independent  $\mbox{Unif}[-\pi/2, \pi/2]$, 
 and  the treatment variable $A \in \{1,2\}$ is  generated independently of  $\bm{X}$ and 
the error term $\epsilon \sim \mathcal{N}(0, 0.5^2)$, 
such that $\mbox{Pr}(A=1) = \mbox{Pr}(A=2) = 1/2$.  
%In the data generating
Under model (\ref{sim3.model}),
 there are  only $2$  true treatment effect-modifiers, $X_1$ and $X_2$. 
 In terms of model (\ref{the.model}), we can write 
 %with
  the $A$-by-$\bm{X}$ interaction effect component functions for model (\ref{sim3.model})  as: $g_{1,A}^{\ast}(X_1) =   (A-1.5)X_1$ and $g_{2,A}^{\ast}(X_2) = 2 (A-1.5) \cos(X_2)$, 
  and $g_{j,A}^{\ast}(X_j) =0$ $(j=3,\ldots,p)$, i.e., 
 the other $p-2$ covariates are ``noise'' covariates, 
%  i.e., we can write $g_{j,A}^{\ast}(X_j) =0$ $(j=3,\ldots,p)$,  
that are not consequential for optimizing ITRs. 
 Also, in (\ref{sim3.model}), there are   $10$ covariates $X_j$ $(j=1,\ldots, 10)$,  among the $p$ covariates, 
 associated with the $\bm{X}$ main effects. % $X_j$ $(j=1,\ldots, p)$.  
Under this setting, the contribution to the variance of the outcome from the $\bm{X}$ main effect component  
was about $2$ times larger than that from the interaction effect component.  

We consider two approaches to  treatment effect-modifier selection: 1) the proposed additive regression approach (\ref{LS5}) 
that specifies a sparse set of functions $\{ g_{j}^\ast, j=1,\ldots, p\}$, 
estimated via  Algorithm~\ref{algorithm1}, with 
the  dimension of the cubic $B$-spline basis  $\bm{\Psi}_j$ in (\ref{eq.5}) set to be $d_j = 6$ $(j=1,\ldots,p)$; 
and 2) the linear regression (MC)  approach of \cite{MC}, 
 \begin{equation} \label{the.mc.approach}
\underset{\{  \beta_j \in \mathbb{R} \}}{\text{minimize}} \quad  \mathbb{E} \bigg[ \big\{ Y -   \sum_{j=1}^p (A - 1.5 ) X_j \beta_j   \big\}^2 \bigg]  + \lambda  \sum_{j=1}^p | \beta_j |, 
\end{equation}
which specifies a sparse vector $\bm{\beta}^\ast = (\beta_1^\ast, \ldots, \beta_p^\ast )^\top \in \mathbb{R}^p$. Given each simulated dataset, the tuning parameter $\lambda >0$ is chosen to minimize a $10$-fold cross-validated prediction error.

\begin{figure} [H] 
\begin{center}
\begin{tabular}{c}
\begin{tabular}{c}  \includegraphics[width=5.7in, height = 1.5in]{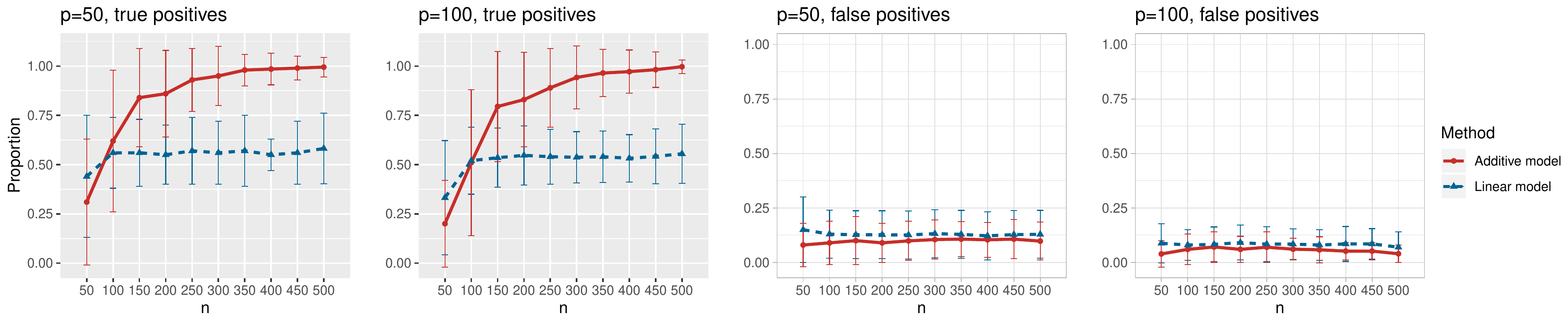} \end{tabular}.  
\end{tabular}
\end{center}
\vspace{-0.3in}
\caption{
The proportion of the relevant covariates (i.e., the treatment effect-modifiers) 
 correctly selected  (the ``true positives''; left two panels),
 and the irrelevant covariates  (i.e., the noise covariates)  incorrectly selected (the ``false positives''; right two panels),  respectively (and $\pm 1$ standard deviation), 
as the training sample size $n$ varies from $50$ to $500$, for each $p \in \{50, 100\}$. Two methods were compared: 1) the proposed additive model (solid curves), and 2) the linear model (MC) approach (dashed curves). 
} \label{fig.sim.result.vs}
\end{figure}

Figure~\ref{fig.sim.result.vs} summarizes the results of the treatment effect-modifier selection performance 
with respect to 
the true/false positive rates (the left/right two panels, respectively), 
comparing 
the proposed additive  regression to the linear regression approach, 
which are reported as the averages (and $\pm 1$ standard deviations) across the $200$ simulation replications.   
 Figure~\ref{fig.sim.result.vs} illustrates that, 
 for the both $p=50$ and $p=100$ cases,  
the proportion of the correctly selecting treatment effect-modifiers 
(i.e., the ``true positive''; the left two panels)
 of the additive regression method (the red solid curves) tends to $1$, with $n$ increasing  from $n=50$ to $n=500$, 
while the proportion of incorrectly selecting treatment effect-modifiers (i.e., the ``false positive''; the right two panels) tends to be bounded above by a small number. 
On the other hand,  
the proportion of correctly selecting  treatment effect-modifiers  
 for 
the linear regression method (the blue dotted curves)   
tends to be only around $0.5$ 
for both choices of $p$. The linear regression method selects only one treatment effect-modifier $X_1$ which has a linear interaction effect with $A$,  
while generally not selecting the other treatment effect-modifier $X_2$, i.e., 
the one that has the nonlinear (cosine) interaction effect (see (\ref{sim3.model}) for the setting) with $A$.

\subsection{Individualized treatment rule estimation performance} \label{sec.sim.ITR}

In this subsection, we assess the optimal ITR estimation 
performance of the proposed method based on  simulations. 
We generate a vector of  covariates $\bm{X} = (X_1, \ldots, X_p)^{\top} \in \mathbb{R}^p$ 
based on  a multivariate normal distribution with each component having 
the marginal distribution $\mathcal{N}(0,  (\pi/2)^2)$ with
the correlation between the components $\mbox{corr}(X_j, X_k) = 0.1^{|j-k|}$. 
In this illustration, we consider $p = 50$. 
 Responses were generated, 
for 1)  ``\textit{nonlinear}'' $A$-by-$\bm{X}$ interactions: 
\begin{equation} \label{sim.model1}
Y =   \  \delta  \sum_{j=1}^5 \sin(X_j)  \  + \  2(A-1.5) \big\{ \cos(X_1) - \cos(X_2)  + \xi \sin(X_1  X_2) \big\}  \ + \ \epsilon \quad  A \in \{1,2\}, 
\end{equation}
and  for 2) ``\textit{linear}'' $A$-by-$\bm{X}$ interactions:   
\begin{equation} \label{sim.model2}
Y =   \  \delta  \sum_{j=1}^5 \sin(X_j)  \  +  \ (A-1.5)  \big\{ X_1   - X_2  + \xi X_1  X_2 \big\}  \ + \  \epsilon \quad  A \in \{1,2\}, 
\end{equation}
where the treatment variable 
$A \in \{1,2\}$ is  generated independently from 
the covariates $\bm{X}$ and 
the error term $\epsilon \sim \mathcal{N}(0, 0.5^2)$, 
such that $\mbox{Pr}(A=1) = \mbox{Pr}(A=2) = 1/2$.  
%in which $A$, $\bm{X}$ and the error term $\epsilon \sim \mathcal{N}(0, 0.5^2)$ are all generated independently.
%independently of  the covariates and the error term $\epsilon \sim \mathcal{N}(0, 0.5^2)$, such that $\mbox{Pr}(A=1) = \mbox{Pr}(A=2) = 1/2$. 
Models   (\ref{sim.model1}) and (\ref{sim.model2})  are indexed by a pair $\{\delta, \xi\}$. 
The parameter  $\delta \in \{1,2\}$ %in models  (\ref{sim.model1}) and (\ref{sim.model2})  
controls the proportion of the variance of the response $Y$ attributable to the $\bm{X}$ ``main'' effect: 
$\delta = 1$ corresponds to a \textit{moderate} $\bm{X}$ main effect contribution; 
 $\delta = 2$ corresponds to a \textit{large} $\bm{X}$ main effect contribution. 
 Estimation of  the interaction effect becomes more difficult with a larger $\delta$. 
 The parameter  $\xi \in \{0,1\}$ % in models  (\ref{sim.model1}) and (\ref{sim.model2})  
 determines whether 
 the $A$-by-$\bm{X}$  interaction effect term 
  has an exact additive regression structure $(\xi = 0)$ or whether 
  it deviates from an additive structure $(\xi = 1)$. In the case of $\xi=0$, the proposed model (\ref{the.model}) is correctly specified, 
  whereas, for the case of  $\xi=1$, model (\ref{the.model}) is misspecified. 
For each scenario, we consider the following four approaches to estimating the optimal ITR $\mathcal{D}^{opt}$ in (\ref{d.opt}).

\begin{enumerate}
\item[1.]
The proposed additive regression approach (\ref{LS5}),   
estimated via  Algorithm~\ref{algorithm1}.  
The  dimension of the basis function  $\bm{\Psi}_j$ in (\ref{eq.5}) is taken to be $d_j = 6$ $(j=1,\ldots,p)$. 
Given  estimates
 $\{ \hat{g}_{j}^\ast \}$, 
the estimate of $\mathcal{D}^{opt}$ in (\ref{d.opt}) is $\hat{\mathcal{D}}^{opt}(\bm{X}) =  \operatorname*{arg\,max}_{a \in \{1,\ldots,L \} }    \sum_{j=1}^p \hat{g}_{j,a}^\ast(X_{j})$.

\item[2.]

The linear regression (MC)  approach  (\ref{the.mc.approach})  of \cite{MC}, 
implemented through the R-package \texttt{glmnet}, 
with the sparsity tuning parameter $\lambda$ selected by minimizing a 10-fold cross-validated prediction error. 
Given an estimate $\hat{\bm{\beta}}^\ast = (\hat{\beta}_1^\ast,\ldots, \hat{\beta}_p^\ast)^\top$, 
the corresponding estimate of $\mathcal{D}^{opt}$ in (\ref{d.opt}) is 
$\hat{\mathcal{D}}^{opt}(\bm{X}) =  \operatorname*{arg\,max}_{a \in \{1,\ldots,L \} }  \sum_{j=1}^p (a - 1.5 ) X_j \hat{\beta}_j^\ast$. 

\item[3.]
The outcome weighted learning (OWL) method \citep[][]{Zhao.2012} based on a Gaussian  radial kernel, implemented in the R-package \texttt{DTRlearn}, 
with a set of feature transformed (FT) covariates  $\{ \hat{g}_{j,1}^\ast(X_{j}), j=1,\ldots,p \}$ used as an input to the OWL method, 
in which the functions $\hat{g}_{j,1}^\ast(\cdot)$ $(j=1,\ldots,p)$ are obtained from the approach in 1.  
To improve the efficiency of the OWL,  we employ  the augmented OWL approach of \cite{Augmented.OWL}. The inverse bandwidth parameter  $\sigma_n^2$ 
and the tuning parameter  $\kappa$ 
in \cite{Zhao.2012} are chosen from the grid of $(0.01, 0.02, 0.04, \ldots, 0.64, 1.28)$ 
and that of $(0.25, 0.5, 1, 2, 4)$ (the default setting of \texttt{DTRlearn}), respectively, based on a $5$ fold cross-validation.

\item[4.]
The same (OWL) approach as in 3 but based on the original features $\{ X_j, j=1,\ldots,p\}$.

\end{enumerate}

 For each simulation run,   
 we estimate $\mathcal{D}^{opt}$ from each of the four methods 
 based on a training set (of size $n \in \{250, 500\}$), and  for evaluation of these methods, we 
compute  the value 
$V(\hat{\mathcal{D}}^{opt})$ in  (\ref{value.eq}) for each estimate  $\hat{\mathcal{D}}^{opt}$,    
using a Monte Carlo approximation 
based on a random sample of size $10^3$. 
Since we know the true data generating model in simulation studies, 
the optimal 
 $\mathcal{D}^{opt}$  can be determined 
for each simulation run. 
Given each estimate $\hat{\mathcal{D}}^{opt}$ of $\mathcal{D}^{opt}$, 
we report  $V(\hat{\mathcal{D}}^{opt})  - V(\mathcal{D}^{opt})$,  
as the performance measure of $\hat{\mathcal{D}}^{opt}$. 
A larger value (i.e., a smaller difference from the optimal value) of the measure indicates  better performance.

\begin{figure} [H]
\begin{center}
\begin{tabular}{c}
\begin{tabular}{c}  \includegraphics[width=5.7in, height = 3.8in]{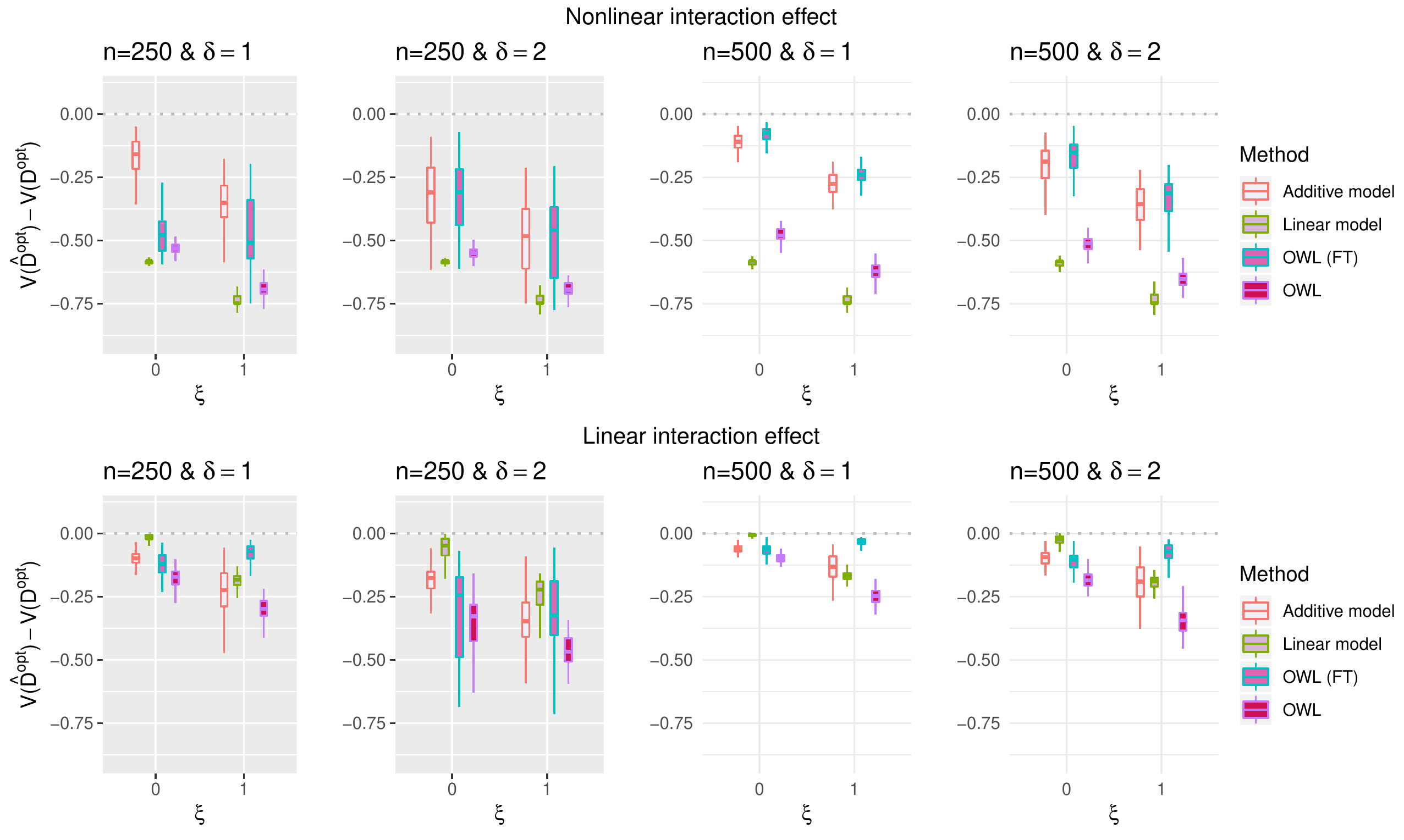} \end{tabular}.  
\end{tabular}
\end{center}
\vspace{-0.3in}
\caption{
Boxplots obtained from  $100$ Monte Carlo simulations 
comparing 4 approaches to estimating $\mathcal{D}^{opt}$, 
given each scenario indexed by $\xi \in \{0,1\}$, $\delta \in \{1,2\}$ and  $n\in \{250, 500\}$, for the \textit{nonlinear} $A$-by-$\bm{X}$ interaction effect case in the top panels, and the \textit{linear} $A$-by-$\bm{X}$ interaction effect  case in the bottom panels. 
The dotted horizontal line represents the optimal value corresponding to $\mathcal{D}^{opt}$.   
} \label{fig.sim.result2}
\end{figure}

In Figure \ref{fig.sim.result2},  
we present the boxplots, obtained from $100$ simulation runs, 
 of the normalized
  values $V(\hat{\mathcal{D}}^{opt})$ 
 (normalized by the optimal values $V(\mathcal{D}^{opt})$)  of the decision rules  $\hat{\mathcal{D}}^{opt}$ 
estimated from the four approaches, 
for each combination of $n \in \{250, 500 \}$, 
$\xi \in \{0, 1\}$ (corresponding to   \textit{correctly-specified} or \textit{mis-specified}  additive interaction effect models, respectively) and 
 $\delta \in \{1, 2 \}$  (corresponding to \textit{moderate} or  \textit{large} main effects, respectively), 
  for the \textit{nonlinear}
 and the  \textit{linear}
  $A$-by-$\bm{X}$ interaction effect scenarios, in the top and bottom panels, respectively. 
The proposed additive regression clearly outperforms the OWL (without feature transformation) method in all scenarios (both the top and bottom panels), and the linear regression approach in all of the \textit{nonlinear} $A$-by-$\bm{X}$ interaction effect scenarios  (the top panels). 
For the  \textit{linear} $A$-by-$\bm{X}$ interaction effect scenarios (the bottom panels), 
when $\xi = 0$ (i.e., when the linear interaction model is correctly specified),  
the linear regression  
outperforms the additive regression, but only slightly, 
whereas 
 if the underlying model deviates from the exact linear structure (i.e., $\xi = 1$ in model (\ref{sim.model2})) 
 and $n=500$, 
 the more flexible additive regression tends to outperform the linear model. 
This suggests that, in the absence of prior knowledge about the form of the interaction effect, for optimizing ITRs, 
 employing the proposed additive regression is more suitable  than the linear regression. 
Comparing the two OWL methods (OWL (FT) and OWL), 
Figure \ref{fig.sim.result2} illustrates that the feature transformation provides a considerable benefit in their performance. 
This suggests  the utility of the estimated component functions $\{\hat{g}_j^\ast, j=1,\ldots,p \}$ of the proposed additive model 
as a potential feature transformation and selection tool for a machine learning algorithm for optimizing ITRs.    
Comparing the cases with $\delta = 2$ to those with $\delta = 1$, 
 the increased magnitude of the main effect generally dampens the performance of all approaches, 
as the ``noise'' variability in the data generation model increases.

\section{Application to data from a depression clinical trial}\label{sec.application}

In this section, we illustrate the utility of the proposed additive regression  for estimating treatment effect-modification and optimizing individualized treatment rules, 
using  data from a depression clinical trial study, comparing an antidepressant and placebo for treating major depressive disorder  \citep{embarc}.  
   The goal of the study is to identify baseline characteristics that are associated with differential response to the antidepressant versus placebo 
   and to use those %pretreatment 
   characteristics to guide treatment decisions when a patient presents for treatment.  
%The study collected baseline patient data, prior to treatment assignment. Selection of potential treatment effect-modifiers of $p=46$ pretreatment covariates was based on their (tier1) influential relationship with MDD clinical outcome as indicated by previous literature. The pretreatment covariates $\bm{X} = (X_1, X_2, \ldots, X_{46})^\top \in \mathbb{R}^{46}$ consist of various functional data such as  magnetic resonance imaging (MRI), functional magnetic resonance imaging (fMRI) and electroencephalogram (EEG) measures that are reduced to scalar values and patient clinical characteristics such as baseline age and week 0 Hamilton Rating Scale for Depression (HRSD) score  \citep{embarc.design2017}. 

 Study participants (a total of $n = 166$ participants)
were randomized to either  placebo  ($A=1$; $n_1 = 88$) or an antidepressant (sertraline) ($A=2$; $n_2 = 78$). Subjects were monitored 
for 8 weeks after initiation of treatment, and 
the primary endpoint of interest was the Hamilton Rating Scale for Depression  (HRSD) score at week 8. 
The outcome $Y$ was taken to be the improvement in symptoms severity 
from baseline  to week $8$, taken as the difference, i.e., we take:  
week 0 HRSD score - week 8 HRSD score.  
(Larger values of the outcome $Y$ are considered desirable.) 
 The study collected baseline patient clinical data, prior to treatment assignment. 
   These pretreatment clinical data 
    $\bm{X} = (X_1, X_2, \ldots, X_{13})^\top  \in \mathbb{R}^{13}$ include:
$X_1=$ Age at evaluation; 
$X_2=$ Severity of depressive symptoms measured by the HRSD at baseline;
$X_3= $ Logarithm of duration (in month) of the current major depressive episode; 
and $X_4=$ Age of onset of the first major depressive episode.
In addition to these standard clinical assessments, patients underwent neuropsychiatric testing at baseline 
to assess psychomotor slowing, working memory, reaction time (RT) 
and cognitive control (e.g., post-error recovery), 
as these behavioral characteristics are believed to correspond to biological phenotypes related to response to antidepressants \citep{embarcDesign}  and are considered as potential modifiers of the treatment effect. 
These neuropsychiatric baseline test measures include:
$X_5=$ (A not B) RT-negative;
$X_6=$ (A not B) RT-non-negative;
$X_{7}= $(A not B) RT-all;
$X_{8}= $ (A not B) RT-total correct; % \citep{AnotB};
$X_9=$ Median choice RT; % \citep[][]{choiceRT};
$X_{10}=$ Word fluency; % \citep[][]{wordFluency};
$X_{11}=$ Flanker accuracy;
$X_{12}=$ Flanker RT;
$X_{13}=$ Post conflict adjustment. % \citep{flanker}.  

\begin{figure} [H]
\begin{center}
\begin{tabular}{c}  \includegraphics[width=5.7in, height = 2in]{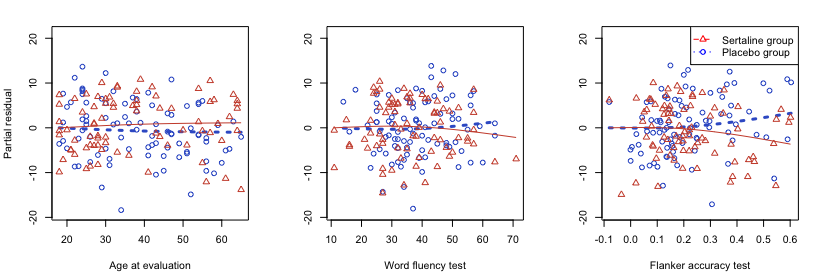} \end{tabular}  
\end{center}
\vspace{-0.2in}
\caption{
Scatterplots of partial residuals vs. the covariates associated with  estimated non-zero component functions $\{ \hat{g}_j^\ast \ne  0  \}$
 %The partial residuals
 for placebo (blue circles) 
and %those for
 the active drug (red triangles) treated participants. 
For each panel, 
the blue dashed curve represents 
$\hat{g}_{j,1}^\ast(\cdot)$, 
 corresponding to the placebo ($a=1$), 
and the red solid curve represents 
  $\hat{g}_{j,2}^\ast(\cdot)$, corresponding to the active drug ($a=2$). 
} \label{fig.embarc.2}
\end{figure}

The proposed approach (\ref{LS5}) to estimating the $A$-by-$\bm{X}$ interaction effect part $\sum_{j=1}^p g_{j,A}^\ast(X_j)$ of  model (\ref{the.model}), estimated via  Algorithm~\ref{algorithm1},
 simultaneously selected  $3$ pretreatment covariates as treatment effect-modifiers: 
 $X_1$ (``Age at evaluation''), $X_{10}$ (``Word fluency test'') and $X_{11}$ (``Flanker accuracy test''). 
% ``flanker accuracy'' (a neuropyschiatric test score); 
%``eeg\_LDAEP\_N1dipole" (regression line for 60-100 dB N1 dipole amplitudes of EEG current source density); 
% ``fmri\_right\_amygdala'' (fMRI emotion recognition task measure of right amygdala); 
%  ``fmri\_leftVS\_br3'' (fMRI measure of left ventral striatum and perigenual cingulate resting coupling); 
% ``dti\_lh\_superiortemporal'' (DTI fractional anisotropy measure of left superior temporal region); 
%``eeg\_net\_alpha'' (right hemisphere  EEG net alpha; closed-minus-open factor means). 
 Figure~\ref{fig.embarc.2} illustrates 
 the estimated non-zero component functions $\{ \hat{g}_j^\ast \ne  0, j=1,\ldots,13  \}$ 
 (i.e., the component functions 
 corresponding to the selected covariates $X_1$, $X_{10}$ and $X_{11}$) 
  and the associated partial residuals. 
%Overall, treatment effect-modification appears to be rather moderate but exhibit some nonlinearities. 
The linear regression approach (\ref{the.mc.approach}) to estimating the $A$-by-$\bm{X}$ interactions selected a single covariate, $X_{11}$, as a treatment effect-modifier.

\begin{figure} [H]
\begin{center}
\begin{tabular}{c}
\begin{tabular}{c}  \includegraphics[width=4.2in, height = 2.2 in]{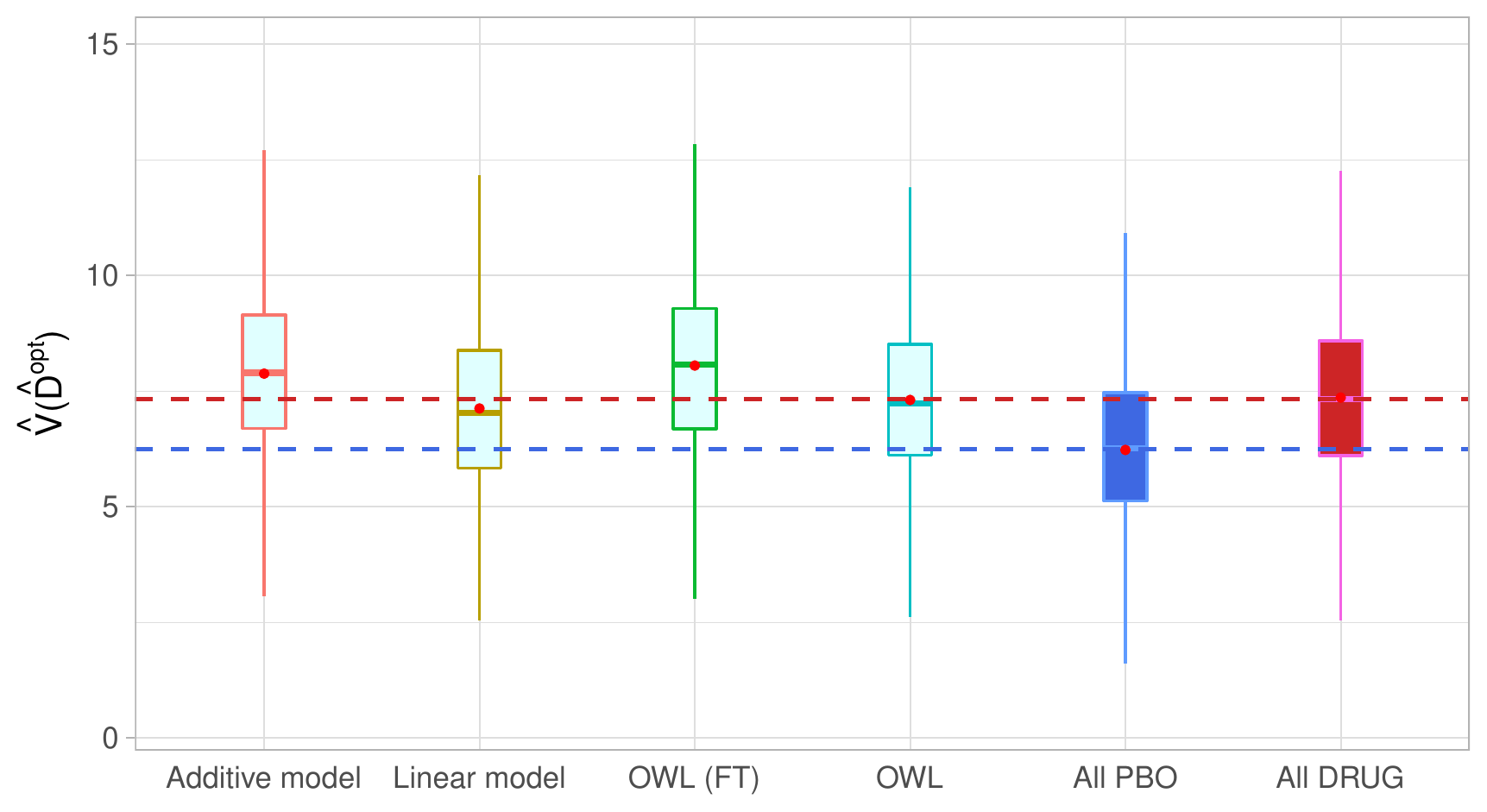} \end{tabular}.  
\end{tabular}
\end{center}
\vspace{-0.3in}
\caption{
Boxplots of the estimated values %(\ref{value.estimator}) 
of the treatment rules $\hat{\mathcal{D}}^{opt}$ estimated from $6$ approaches, obtained from $500$ randomly split testing sets.  
Higher values are preferred. 
} \label{fig.embarc.4}
\end{figure}

To evaluate the performance of the ITRs ($\hat{\mathcal{D}}^{opt}$) obtained from the four different approaches described in Section~\ref{sec.sim.ITR}, 
we randomly split the data into a training set  and a testing set  (of size $\tilde{n})$ using a ratio of $5$ to $1$, replicated $500$ times, 
each time computing an ITR $\hat{\mathcal{D}}^{opt}$ 
 based on the training set, then estimating its value $V(\hat{\mathcal{D}}^{opt})$ in (\ref{value.eq})
by an inverse probability weighted estimator \citep{Murphy2005}: 
%\begin{equation}  \label{value.estimator}
$\hat{V}(\hat{\mathcal{D}}^{opt}) = \sum_{i=1}^{\tilde{n}} Y_{i} I_{(A_i = \hat{\mathcal{D}}^{opt}(\bm{X}_i) )} / \sum_{i=1}^{\tilde{n}} I_{(A_i =\hat{\mathcal{D}}^{opt}(\bm{X}_i)) }$, 
%\end{equation}
 computed based on  the testing set of size $\tilde{n}$. 
% Figure~\ref{fig.embarc.4} illustrates the resulting boxplots. 
% The resulting boxplots are illustrated in Figure~\ref{fig.embarc.4}. 
  For comparison, 
we also include two na\"{\i}ve %treatment decision 
rules: 
 treating all patients with placebo (``All PBO'')  and treating all patients with the active drug (``All DRUG''), each regardless of the individual patient's characteristics $\bm{X}$.   The resulting boxplots obtained from the $500$ random splits are illustrated in Figure~\ref{fig.embarc.4}.  A larger value of the measure indicates  better performance.

The results in Figure~\ref{fig.embarc.4} demonstrate that the proposed additive regression approach, % to optimizing ITRs, 
which allows nonlinear flexibility in developing ITRs, 
 tends to outperform the linear regression approach,  in terms of the estimated value. % $\hat{V}(\hat{\mathcal{D}}^{opt})$. % (\ref{value.estimator}). 
The additive regression approach also shows some superiority over the method OWL (without feature transformation).  
In comparison to the OWL methods, the proposed additive regression, in addition to its superior computational efficiency, 
 provides a means of  simultaneously selecting treatment effect-modifiers 
and allows a visualization for the heterogeneous effects attributable to each estimated treatment effect-modifier as in Figure~\ref{fig.embarc.2}, 
which is an appealing feature in practice. 
Moreover, the estimated component functions $\{\hat{g}_j^\ast, j=1,\ldots,p\}$ of the proposed regression 
 provide an effective means of performing feature transformation for $\{X_j, j=1,\ldots,p\}$. As in Section~\ref{sec.sim.ITR}, 
the feature transformed OWL approach appears to have a considerable improvement over the OWL that bases on the original untransformed  covariates.

\section{Discussion} \label{sec.conclusion}

In this paper, we have developed a sparse additive model, via a structural constraint, specifically geared to identify and model treatment effect-modifiers.
%\st{we have shown how we can modify sparse additive models by imposing an appropriate structural constraint  to perform variable selection specifically  for treatment effect-modifiers.}  
The approach utilizes an efficient back-fitting algorithm for model estimation and variable selection. % \st{to estimate the proposed model. }
The proposed sparse additive model for treatment effect-modification extends existing linear model-based regression methods by providing nonlinear flexibility to modeling treatment-by-covariate interactions. 
 Encouraged by our simulation results and the application,  future work will investigate the asymptotic properties related to treatment effect-modifier selection and estimation consistency, in addition to developing hypothesis testing procedures for treatment-by-covariates interaction effects. %\st{based on the proposed models.}

Modern advances in biotechnology, using measures of brain structure and function obtained from neuroimaging modalities (e.g., MRI, fMRI and EEG), show the promise of discovering potential biomarkers for heterogeneous treatment effects.
%\st{often involve complex patient information such as brain structure  and function, measured from neuroimaging modalities such as MRI, fMRI and EEG.} 
%\st{This type of} 
These high dimensional data modalities are often in the form of curves or images and can be viewed as functional data \citep[e.g.,][]{fda}. Future work will also extend the additive model approach to the context of 
functional additive regression \citep[e.g.,][]{FAR, FRAME}.  The goal of these extensions will be to handle a large number of functional-valued 
covariates while achieving simultaneous variable selection, 
which will extend current functional linear model-based methods for precision medicine \citep{McKeague.Qian.2014,Ciarleglio.scalar.on.functions, Ciarleglio.jrss.c} 
to a more flexible functional regression 
setting.

\section{Software}
\label{sec5}

\texttt{R}-package \texttt{samTEMsel} (Sparse Additive Models for Treatment Effect-Modifier Selection) contains \texttt{R}-codes to perform the methods proposed in the article, and is publicly available on \texttt{GitHub} (\texttt{syhyunpark/samTEMsel}).

\section{Supplementary Materials}
\label{sec6}

The Supplementary Materials  including the proof of Theorem~\ref{theorem1} and the \texttt{R}-codes and the dataset used in this article are available online at \url{http://biostatistics.oxfordjournals.org}.

\section*{Acknowledgments}

This work was supported by National Institute of Health (NIH) grant 5 R01 MH099003.

{\it Conflict of Interest}: None declared.

%\newpage

\bibliographystyle{biorefs}
\bibliography{refs}

%\newpage

%\appendix

\newpage

\section*{Supplementary Materials}
\subsection*{Proof of Theorem 1}

%\subsubsection*{Proof of Theorem 1} 
\begin{proof}
The squared error criterion on the right-hand side of (2.6) of the main manuscript %(\ref{LS4}) 
is 
\begin{equation} \label{LS.framework}
\begin{aligned}
 \mathbb{E} \bigg[ \big\{ Y -    \sum_{j=1}^p g_{j,A} ( X_j)  
 \big\}^2  \bigg]  
 \propto  & \ \mathbb{E} \bigg[ Y  \sum_{j=1}^p g_{j,A}(X_j)  
 -\big\{ \sum_{j=1}^p g_{j,A}(X_j) 
  \big\}^2 /2
  \bigg] \quad  \mbox{(with respect to } \{g_{j}\} ) 
  \\
 =  & \ \mathbb{E} \bigg[ \big\{ \mu^\ast(\bm{X}) +   \sum_{j=1}^p g_{j,A}^\ast(X_j) 
 \big\}
 \sum_{j=1}^p g_{j,A}(X_j)  
- \big\{ \sum_{j=1}^p g_{j,A}(X_j)  
 \big\}^2 /2
  \bigg]  \\
 = & \ \mathbb{E} \bigg[ \mu^\ast(\bm{X})  \sum_{j=1}^p g_{j,A}(X_j) \bigg] +   \mathbb{E}\bigg[  \big\{ \sum_{j=1}^p g_{j,A}^\ast(X_j)  
  \big\} \big\{ \sum_{j=1}^p g_{j,A}(X_j)  
   \big\}   
  -\big\{ \sum_{j=1}^p g_{j,A}(X_j)  
  \big\}^2 /2   \bigg]  \\
 = & \  \mathbb{E}\bigg[  \big\{ \sum_{j=1}^p g_{j,A}^\ast(X_j)  
  \big\} \big\{ \sum_{j=1}^p g_{j,A}(X_j)  
   \big\}   
  -\big\{ \sum_{j=1}^p g_{j,A}(X_j)  
  \big\}^2 /2   \bigg], 
\end{aligned} 
\end{equation}
where the last equality follows from the constraint $ \mathbb{E}[g_{j,A}( X_j) | X_j ]  = 0$ $(j=1,\ldots,p)$ in (2.6) of the main manuscript imposed on $\{g_j\}$, 
and by noting 
$\mathbb{E} \big[  \mu^\ast( \bm{X} ) \sum_{j=1}^p g_{j,A} \big( X_j \big)  \big]
=
\mathbb{E} \big[  \mathbb{E} \big[ \mu^\ast( \bm{X} ) \sum_{j=1}^p g_{j,A}(X_j) | \bm{X} \big]  \big]
= 
\mathbb{E} \big[ \mu^\ast(\bm{X} ) \sum_{j=1}^p   \mathbb{E} \big[  g_{j,A}(X_j)   | X_j \big]\big] =0$.  
From (\ref{LS.framework}), the squared error criterion in (2.6) of the main manuscript can be expressed as: 
\begin{equation} \label{LS.framework2}
\underset{ \{ g_{j} \in \mathcal{H}_j \} }{\text{argmin}} \ \mathbb{E}\bigg[ \big(Y -    \sum_{j=1}^p g_{j,A} \big( X_j \big)  \big)^2  \bigg] 
 \ =  \ 
  \underset{ \{ g_{j} \in \mathcal{H}_j \} }{\text{argmin}} \ \mathbb{E} \bigg[ \big( \sum_{j=1}^p g_{j,A}^\ast\big( X_j \big)   -    \sum_{j=1}^p g_{j,A} \big( X_j \big)  \big)^2  \bigg].  
  \end{equation} 
In the following, we closely follow the proof of Theorem 1 in \cite{SAM}. 
The constrained objective function in (2.6) of the main manuscript  can be rewritten in Lagrangian form as: 
%The Lagrangian in (\ref{LS4}) can be rewritten as: 
\begin{equation}\label{H}
Q( \{g_j\}; \lambda)  := 
 \mathbb{E} \bigg[ \big( \sum_{j=1}^p g_{j,A}^\ast(X_j)   -    \sum_{j=1}^p g_{j,A}(X_j)  \big)^2  \bigg]
+ \lambda  \sum_{j=1}^p \lVert g_{j} \rVert  
\end{equation}
For the notational simplicity, let us write 
$g_j = g_{j,A}(X_j)$. 
 For each $j$, 
consider the minimization of (\ref{H}) 
with respect to the component function $g_{j} \in \mathcal{H}_j$, 
holding the other component functions 
$\{ g_{j'}, j' \ne j\}$ fixed. 
The stationary condition is obtained by setting its Fr\'{e}chet derivative to 0. 
Denote by 
 $\partial_{j} Q( \{g_j\}; \lambda; \eta_{j})$ 
 the directional derivative with respect to $g_{j}$ $(j=1,\ldots,p)$ 
 in the direction, say, $\eta_{j} \in \mathcal{H}_j$. 
 Then, the stationary point of the Lagrangian (\ref{H}) can be formulated as:
\begin{equation}\label{stationary.cond}
 \partial_{j}  Q( \{g_j\}; \lambda; \eta_{j}) 
 = 2 \mathbb{E}\left[  (g_{j} - \tilde{R}_{j} + \lambda \nu_{j}  ) \eta_{j}  \right] =0,  
\end{equation}
 where 
\begin{equation} \label{R.j}
\tilde{R}_{j}  := \sum_{j=1}^p g_{j,A}^\ast(X_j) - \sum_{j' \ne j} g_{j^\prime,A}(X_j)
 \end{equation} 
 is the partial residual for $g_{j}$, and 
 $\nu_{j}$ 
 is an element of the subgradient  
 $\partial  \lVert g_{j} \rVert$, 
 which satisfies 
 $\nu_{j} =   g_{j} /  \lVert g_{j} \rVert $ if $\lVert g_{j} \rVert  \ne 0$, 
 and 
  $\nu_{j} \in \{ s \in \mathcal{H}_j \mid   \lVert s \rVert  \le 1 \} $, otherwise. 
 Using iterated expectations conditional on $X_j$ and $A$, 
 (\ref{stationary.cond}) can be rewritten as 
\begin{equation} \label{stationary.cond2}
  2 \mathbb{E}\left[ \left(g_{j} -  \mathbb{E}\left[ \tilde{R}_{j} | X_j, A \right]  + \lambda \nu_{j}   \right) \eta_{j}   \right] =0.  
\end{equation} 
Since 
 $ g_{j}   -  \mathbb{E}\left[ \tilde{R}_{j}  | X_j, A \right]  + \lambda \nu_{j}   \in \mathcal{H}_j$, 
 we can evaluate (\ref{stationary.cond}) (i.e., (\ref{stationary.cond2})) in the direction: 
$\eta_{j}  =  g_{j}   -  \mathbb{E}\left[ \tilde{R}_{j}  | X_j, A \right]  + \lambda \nu_{j} $, 
implying 
$ \mathbb{E}\left[ \left( g_{j}  -  \mathbb{E}\left[ \tilde{R}_{j}  | X_j, A  \right]  + \lambda \nu_{j}  \right)^2 \right] =0.$ 
This implies: 
\begin{equation} \label{stationary.condition}
g_{j}  + \lambda \nu_{j} =  \mathbb{E}\left[ \tilde{R}_{j} | X_j, A \right]    \quad 
 \mbox{(almost surely).}
\end{equation} 
Let $f_{j}$ denote  
the right-hand side of (\ref{stationary.condition}), i.e., 
$f_{j} (= f_{j,A}(X_j)) := \mathbb{E}\left[ \tilde{R}_{j}  | X_j, A \right]$. 
If $\lVert g_{j} \rVert  \ne 0$, 
then $\nu_{j}  = g_{j}/  \lVert g_{j} \rVert$.  
Therefore, by (\ref{stationary.condition}), we have 
$\lVert f_{j}  \rVert = \lVert  g_{j}+ \lambda  g_{j} /  \lVert g_{j} \rVert    \rVert 
= \lVert  g_{j}  \rVert + \lambda 
\ge  \lambda$.    
On the other hand, 
if $\lVert g_{j} \rVert  = 0$,  
then $g_{j} = 0$ (almost surely) 
and $\lVert \nu_{j} \rVert  \le 1$ which, together with condition (\ref{stationary.condition}), implies that 
$\lVert f_{j} \rVert  \le \lambda$. 
This gives us the equivalence between 
$\lVert f_{j} \rVert  \le \lambda$ and the statement 
$g_{j}= 0$ (almost surely). 
Therefore, 
condition (\ref{stationary.condition}) leads to the following expression: 
\begin{equation*} \label{g.jt.solution}
\left( 1 + \lambda /  \lVert  g_{j}  \rVert  \right)  g_{j}   =  f_{j}  \quad 
\mbox{(almost surely)}
\end{equation*} 
if $\lVert f_{j} \rVert  > \lambda$; otherwise, 
and $ g_{j}  = 0$ (almost surely). This gives the soft thresholding update rule for $g_{j}$. 

The underlying model (2.1) of the main manuscript %(\ref{the.model}) 
indicates that  
 $ \sum_{j=1}^p g_{j,A}^\ast(X_j)  = \mathbb{E}[Y | X, A] - \mu^\ast(\bm{X})$. Thus, 
 (\ref{R.j}) can be equivalently written as: 
 $\tilde{R}_{j} =  \mathbb{E}[Y | X, A] - \mu^\ast(\bm{X})  - \sum_{j' \ne j} g_{j',A}(X_{j^\prime}).$
Therefore, the function $f_{j,A}(X_j) = \mathbb{E}\left[ \tilde{R}_{j}  | X_j, A \right]$ can be written by: 
\begin{equation*} \label{final.eq}
\begin{aligned}
 f_{j,A}(X_j)  \ &=   \mathbb{E}\big[ \mathbb{E}[Y | X, A] - \mu^\ast(\bm{X})  - \sum_{j' \ne j} g_{j', A}(X_{j^\prime}) \mid  X_j, A \big]  \\
 &= \mathbb{E}\big[ \mathbb{E}[Y | X, A]  - \sum_{j' \ne j} g_{j', A}(X_{j^\prime}) |  X_j, A \big]   - \mathbb{E}\big[ \mu^\ast(\bm{X}) |  X_j, A \big] \\
 &= \mathbb{E}\big[ Y - \sum_{j' \ne j} g_{j', A}(X_{j^\prime}) |  X_j, A \big]   - \mathbb{E}\big[\mu^\ast(\bm{X}) |  X_j \big]  \\
 &=  \mathbb{E}\big[ Y - \sum_{j' \ne j} g_{j', A}(X_{j^\prime}) |  X_j, A \big]    - \mathbb{E}\big[\mu^\ast(\bm{X}) + \sum_{j=1}^p g_{j,A}^\ast\big( X_j \big) |  X_j\big]   \\
     &= \mathbb{E}\big[ Y - \sum_{j' \ne j} g_{j', A}(X_{j^\prime}) |  X_j, A \big]   - \mathbb{E}\big[Y  |  X_j \big] \\
     &= \mathbb{E}\big[ Y - \sum_{j' \ne j} g_{j', A}(X_{j^\prime}) |  X_j, A \big]   - \mathbb{E}\big[Y  - \sum_{j' \ne j} g_{j', A}(X_{j^\prime}) |  X_j \big] \\ 
     &= \mathbb{E}\big[ R_{j} |  X_j, A \big]   - \mathbb{E}\big[R_{j}  |  X_j\big],  
\end{aligned} 
 \end{equation*}
 where the fourth equality follows from the identifiability constraint (2.2) of the underlying model (2.1) of the main manuscript, 
 and the sixth equality follows from the  optimization constraint 
 $ \mathbb{E}[g_{j,A}( X_j) | X_j ]  = 0$ $(j=1,\ldots,p)$ in (2.6) of the main manuscript  imposed on $\{g_j\}$; 
this gives the desired expression (3.8) of the main manuscript. %(\ref{proj.1}). 

\end{proof}

\end{document}